# Magnetism and electron spin resonance in single crystalline $\beta$-AgNpO$_2$(SeO$_3$)

E. Jobiliong, Y. Oshima, J.S. Brooks
*Physics/NHMFL, Florida State University, Tallahassee FL 32310*

T.E. Albrecht-Schmitt
*Department of Chemistry and Leach Nuclear Science Center, Auburn University, Auburn, AL 36849*

We report magnetization, susceptibility, electrical transport, and electron spin resonance (ESR) studies of single crystals of $\beta$-AgNpO$_2$(SeO$_3$). Here the valence of the Np sites is expected to be Np(V). We observe a magnetic transition below 8 K, where the transition temperature is dependent on the effective magnetic moment. Although the transition appears to be ferromagnetic, no hysteresis is seen in the magnetization, and the saturation moment above 0.1 T is found to be about 60% of the free NpO$_2$ ion moment. The decrease in the Np moments determined experimentally is thought to arise from crystal field and spin-orbit effects. Although Np(V) is expected to be ESR silent, we observe temperature dependent ESR spectra at ~44 GHz (for fields above the saturation field) that show slight shifts in the g-factor and line width at low temperatures. Our results provide evidence that both Np(V) and Np(IV) valences are present, where the latter may be a minority population. The crystals, although dark in appearance, are electrically insulating ($\rho > 10^{10}$ Ohm-cm) at room temperature.

## 1. Introduction

The study of new compounds containing actinides, especially Neptunium (Np), has been an active area of research for many years [1-10]. The physical properties, and in particular the magnetic properties, are of interest since they depend significantly on the material surrounding the Np site: for instance Np(NpO$_2$)$_2$(SeO$_3$)$_3$ [3], NpO$_2$(O$_2$CH)(H$_2$O) [4], [(NpO$_2$)$_2$(O$_2$C)$_2$C$_6$H$_4$]•6H$_2$O [5], and [Na$_4$(NpO$_2$)$_2$(C$_{12}$O$_{12}$)]•8H$_2$O all utilize oxo donors, however they contain distinct types of ligands coordinating the Np centers [11]. Magnetic transitions have been observed in some materials such as [Na$_4$(NpO$_2$)$_2$(C$_{12}$O$_{12}$)]• 8H$_2$O at 10 K, but the details of the magnetic properties are not well understood. The highest magnetic ordering temperature observed is NpO$_2$(O$_2$CH)(H$_2$O), which is around 12.8 K. The magnetic behavior of NpO$_2$(O$_2$CH)(H$_2$O) is similar to that of (NpO$_2$)$_2$C$_2$O$_4$•4H$_2$O [12], but the interpretation is quite different, since the transitions have been reported to be metamagnetic and ferromagnetic in the two materials, respectively. The magnetic properties of [(NpO$_2$)$_2$(O$_2$C)$_2$C$_6$H$_4$]•6H$_2$O have a complex magnetic structure at low temperature that has been interpreted in terms of concurrent ferromagnetic ordering and metamagnetic behavior.

In the present work we consider the magnetic properties of single crystals of $\beta$-AgNpO$_2$(SeO$_3$). This compound contains neptunyl(V)-neptunyl(V) coordination, so-called cation-cation interactions, whereby the normally terminal oxo atoms of the NpO$_2^+$ cations are instead used to coordinate adjacent NpO$_2^+$ moieties yielding Np(V)···Np(V) distances of approximately 4.2 Å [6]. $\beta$-AgNpO$_2$(SeO$_3$) forms a three-dimensional network composed of sheets of neptunium oxide formed via the cation-cation interactions that are joined together by selenite anions creating small channels that run though the lattice. Ag$^+$ cations fill the channels, as shown in Fig.1. The outer electron configuration of Np(V) is $5f^2$. The magnetic properties of $f$ electron systems are significantly different from $d$-electron systems due to the strong spin-orbit coupling [13].

Moreover, the $5f$ electron systems have a greater exposure to their chemical environments than do the $4f$ electron systems, and hence the crystal field effect plays a significant role with the spin-orbit coupling. Thus, it is necessary to consider the crystal structure, spin-orbit coupling, and crystal-field splitting in order to fully understand the magnetic properties of these compounds.

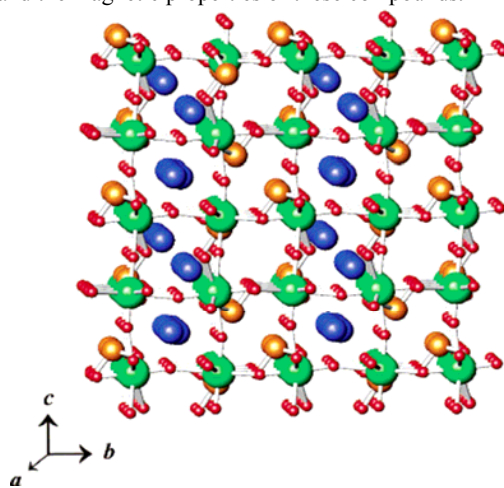

Figure 1. View down the $a$-axis of the three-dimensional structure of $\beta$-AgNpO$_2$(SeO$_3$). Green = Np, blue = Ag, orange = Se, and red = O.

## 2. Experimental

Single crystals of $\beta$-AgNpO$_2$(SeO$_3$) were prepared by methods previously described [6]. The crystals are dark in appearance, and were ~6.6×10$^{-3}$ mm$^3$, ~3.1×10$^{-3}$ mm$^3$, ~2.3×10$^{-3}$ mm$^3$, and ~2.1×10$^{-3}$ mm$^3$ in size for samples 1, 2, 3, and 4 respectively. The magnetization studies were carried out in a Superconducting Quantum Interference Device (SQUID) magnetometer over the temperature range 2 – 200 K, and hysteresis measurements were carried out in the field range -2 T to 2 T. Due to the small size of the samples, the crystals were attached to the surface of a thin quartz pipe with GE varnish. The ESR measurements were performed using the cavity perturbation technique with a millimeter wave vector network analyzer (MVNA) [14]. The MVNA includes a tunable microwave source that covers the frequency range of 8-350 GHz, and a high sensitivity detector. The sample was set on the end-plate of the cavity, so that the sample is always in the oscillatory magnetic field. This is the usual configuration for ESR measurements. An attempt to determine the sample resistance at room temperature was carried out using an electrometer in a two-terminal microprobe configuration.

## 3. Magnetization and susceptibility of $\beta$-AgNpO$_2$(SeO$_3$)

The total magnetization of the first sample studied (sample 1) revealed magnetic ordering below 7 K, as shown in Fig. 2(a). Here, the field was applied at $40^0$ to the [001]. The inverse susceptibility ($1/\chi \equiv B/M$) vs. temperature is shown in the inset of Fig. 2(a). Here $1/\chi$ shows a slow decrease, followed by a faster decrease as $T_C$ is approached from the paramagnetic state. The two slopes most likely arise from two different paramagnetic phases. In Fig. 2(b), we show the total magnetization versus field at 4.2 K. The saturated magnetic moment is $\sim 0.8\mu_B$ and no hysteresis is observed. In the zero field-cooled and field-cooled magnetization shown in Fig. 2(c) we find little difference, reaffirming the absence of hysteresis.

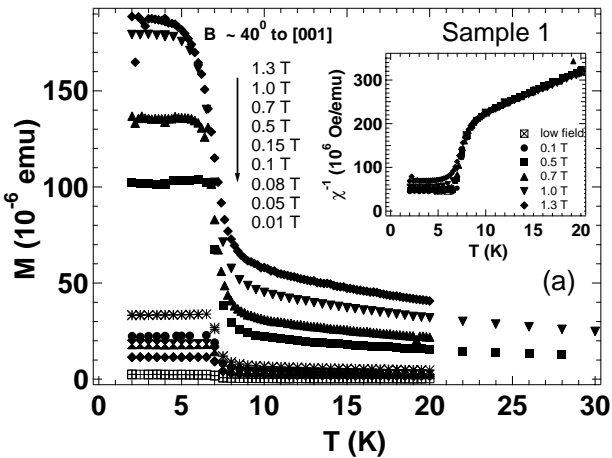

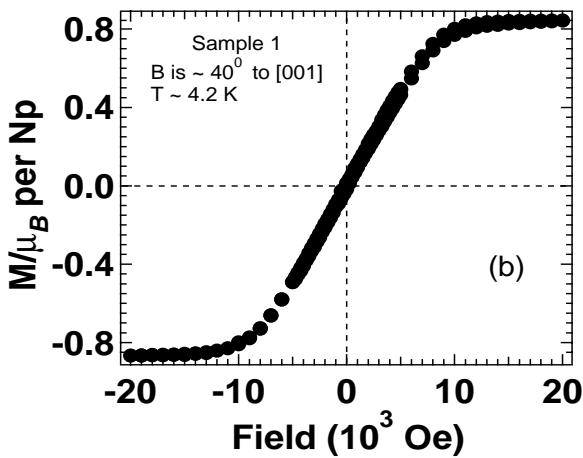

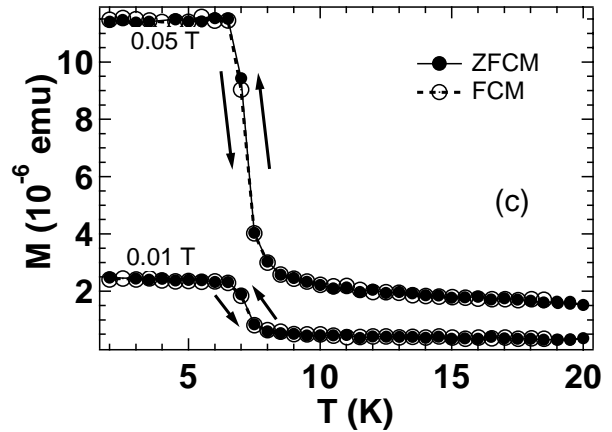

Figure 2. Magnetic properties of $\beta$-AgNpO$_2$(SeO$_3$) for sample 1: (a) The total magnetization vs. temperature for different fields. Inset: Inverse susceptibility vs. temperature. (b) Magnetization per Np (in $\mu_B$) vs. field. (c) Magnetization vs. temperature in ZFCM: zero-field cooled magnetization and FCM: field cooled magnetization.

Fig. 3(a) shows the magnetization versus temperature of sample 2 for the field applied parallel to the [001]. Magnetic ordering appears below 8 K. The inverse susceptibility vs. temperature in the inset of Fig. 3(b) shows a transition from paramagnetic to ferromagnetic behavior. We have used the Curie-Weiss law

$$\chi = \frac{C}{T - T_C} + \chi_0 \quad \ldots\ldots\ldots\ldots\ldots (1)$$

where $C$, $T_C$, and $\chi_0$ are Curie constant, Curie temperature, and temperature-independent susceptibility, respectively to fit the data in the paramagnetic state above $T_C$. We find the parameters corresponding to this fit are $C = 0.672$ K, $T_C = 7.43$ K and $\chi_0 = 0.009$ emu/mol. The effective magnetic moment ($\mu_{eff}$) in the paramagnetic region can be obtained by using

$$C = \frac{N\mu_{eff}^2}{3k_B} \quad \ldots\ldots\ldots\ldots\ldots (2)$$

where $N$ is Avogadro's number and $k_B$ is Boltzmann's constant, which yields $\mu_{eff} = 2.32\mu_B$. The magnetization of sample 2 at T ~ 4.2 K (below $T_C$) is shown in Fig. 2(b). As with sample 1, no hysteresis is observed in sample 2.

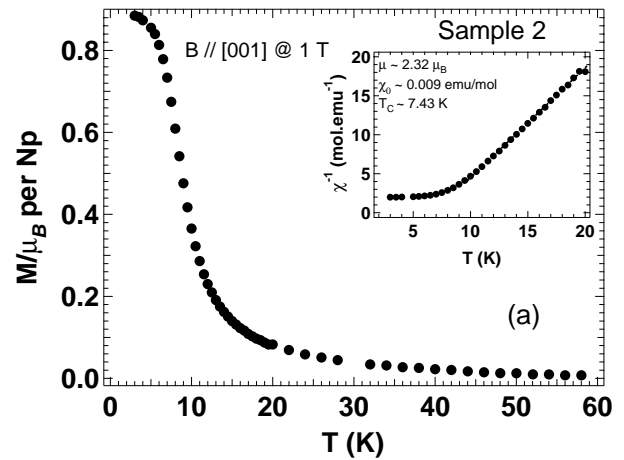

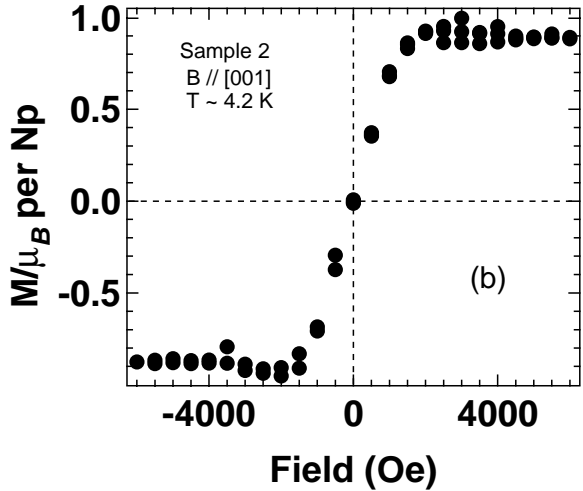

Figure 3. Magnetic properties of $\beta$-AgNpO$_2$(SeO$_3$) for sample 2: (a) The magnetization per Np (in $\mu_B$) vs. temperature. Inset: Inverse susceptibility vs. temperature. (b) Magnetization per Np (in $\mu_B$) vs. field.

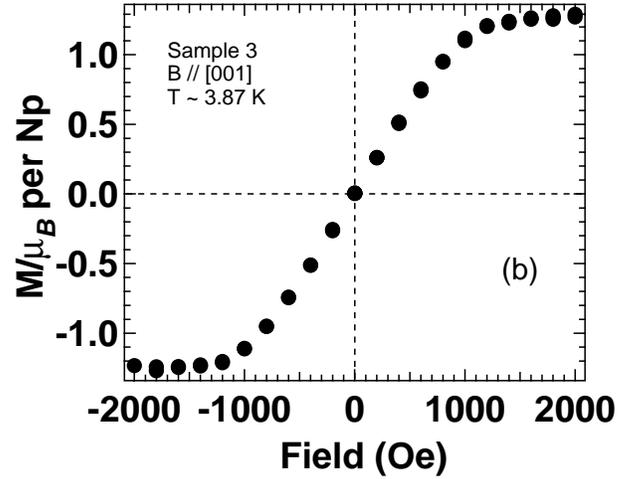

Figure 4. Magnetic properties of $\beta$-AgNpO$_2$(SeO$_3$) for sample 3: (a) The magnetization per Np (in $\mu_B$) vs. temperature. Inset: Inverse susceptibility vs. temperature (b) Magnetization per Np (in $\mu_B$) vs. field.

The magnetization vs. field and temperature for different field orientations for two other samples is shown in Fig. 4 and 5 for sample 3 and 4, respectively. Samples 2, 3 and 4 exhibit similar behavior, and indicate a paramagnetic to ferromagnetic transition at the Curie temperature $T_C$. Although sample 1 also has ferromagnetic order, two different paramagnetic states are apparent above $T_C$. This may be due to the interaction between the local moments of Np ion, which is angle dependent. The angle dependence of the interaction also has been observed in sample 4, which shows a slightly different Curie temperature for the two different orientations. The Curie temperature, effective magnetic moment, saturation moment, and temperature-independent susceptibility obtained from the analysis are summarized for all samples in table 1.

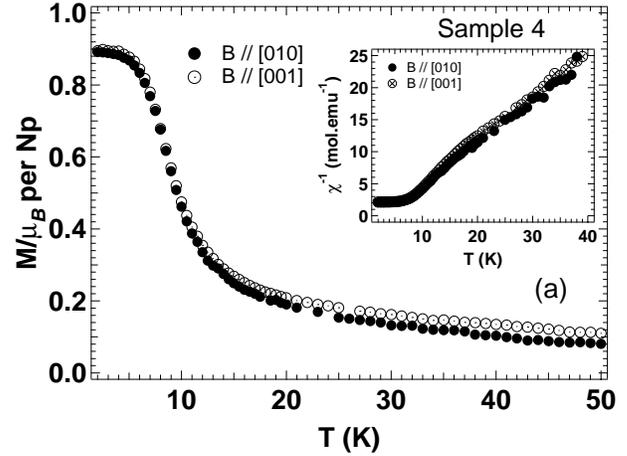

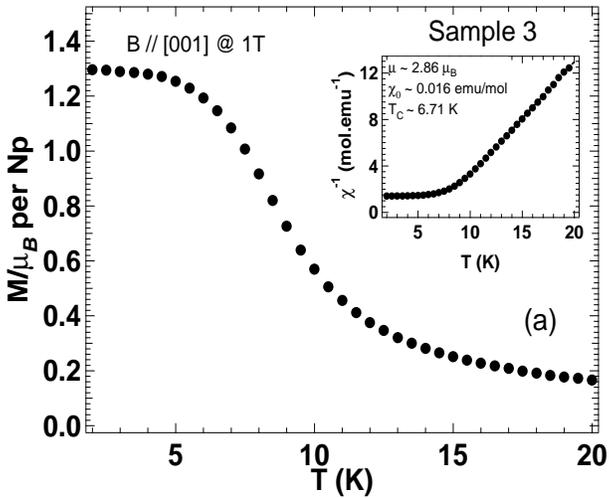

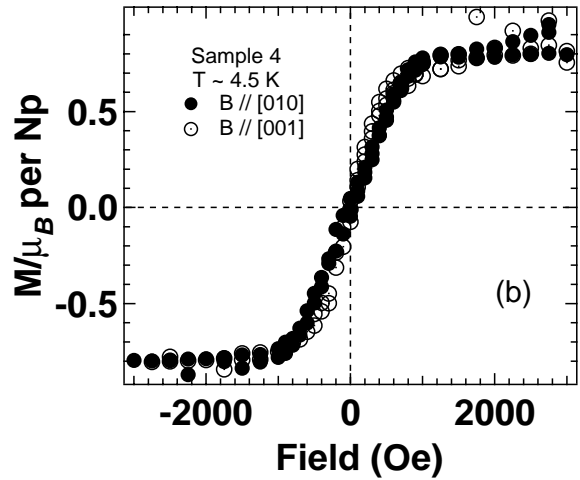

Figure 5. Magnetic properties of $\beta$-AgNpO$_2$(SeO$_3$) for sample 4: (a) The magnetization per Np (in $\mu_B$) vs. temperature. Inset: Inverse susceptibility vs. temperature. (b) Magnetization per Np (in $\mu_B$) vs. field for different orientation.

| Sample | Direction | $\mu_{eff} / \mu_B$ | $\mu_{sat} / \mu_B$ | $T_C$ (K) | $\chi_0$ (emu/mol) |
|---|---|---|---|---|---|
| Sample 1 | B is ~ 40° to [001] | - | 0.82 | 6.2 | - |
| Sample 2 | B is // [001] | 2.32 | 0.89 | 7.43 | 0.009 |
| Sample 3 | B is // [001] | 2.86 | 1.29 | 6.71 | 0.016 |
| Sample 4 | B is // [001] | 3.19 | 0.81 | 4.53 | 0.0191 |
| | B is // [010] | 3.17 | 0.86 | 5.43 | 0.0365 |

Table 1. Field direction, effective and saturation magnetic moments, Curie temperature and temperature independent susceptibility for the single crystals of $\beta$-AgNpO$_2$(SeO$_3$) used in this work.

In the paramagnetic state above $T_C$, a comparison between the experimental magnetic moment and the free ion moment can be made using the spin-orbit coupling (Russell-Saunders coupling) relation

$$\mu_{eff}^T = g_J \sqrt{J(J+1)} \mu_B \quad \ldots\ldots\ldots (3)$$

where $\mu_B$ is Bohr-magneton and

$$g_J = \frac{3}{2} + \frac{S(S+1) - L(L+1)}{2J(J+1)} \quad \ldots\ldots\ldots (4)$$

is the Landé g-value. For Np(V) (S=1; L=5; J=4) $\mu_{eff}^T$ = 3.58 $\mu_B$, and for Np(IV) (S=3/2; L=6; J=9/2) $\mu_{eff}^T$ = 3.62 $\mu_B$. We find the experimental $\mu_{eff}$ is always less than that of either free ion valence. This is due to the effect of the electric field from neighboring atoms in the crystal, i.e. the crystal-field effect. The splitting due to large crystal-field in the 5$f$ system can be in the order 600 K [15]. The splitting will reduce both the ground state and the corresponding effective moment. For instance, an ion of Np(V) in a cubic environment has a $\Gamma_5$ triplet ground state, with $\mu_{eff}$ = 2.8 $\mu_B$ [3]. This reduced value is consistent with the values in Table 1, although the environment in the case of the title material is triclinic.

Below the ordering temperature $T_C$, the experimental saturation moments $\mu_{sat}$ obtained from the magnetization, shown in Fig. 2(b) to 5(b) and in Table 1, are lower than the theoretical values. A reduction in the saturation moment can be significant for heavier ions where there is a large spin-orbit splitting. Here the crystal-field effect and spin-orbit interaction can be comparable. The effect of spin-orbit coupling on the saturation moment can be calculated from $\mu_{sat} = g_J \mu_B J$, and is equal to 3.20$\mu_B$ per atom for Np(V). This estimate of $\mu_{sat}$ does not include the crystal-field effect, which in the solid (especially for the heavy ions) will reduce the value further.

The absence of the hysteresis in the ferromagnetic region implies no magnetic domains are present. For materials with a triclinic crystal structure, the symmetry is broken. Qualitatively, in the presence of the crystal-field, the Np(V) ion in a triclinic environment has only a singlet ground state. This is consistent with Kramers theorem [16], because the Np(V) ion is 5$f^2$, which has an even number of electrons, and no degeneracy. Since the ground state is singlet, there can be no magnetic domain structure, and therefore no hysteresis. In contrast, for Np(IV), the electron configuration is [Rn]5$f^3$, which can be categorized as a Kramers ion. In this case the ground state is a doublet, even in a triclinic environment, where domains and hysteresis will be present. In light of the above, the valence of Np in this compound, based on the bulk magnetization studies, must be predominantly Np(V).

We have investigated the electrical transport of a single crystal of $\beta$-AgNpO$_2$(SeO$_3$) at room temperature and have found the resistivity of the material to be higher than $10^{10}$ $\Omega$.cm. This implies localized electronic states, so the moments must also be localized. The interaction between these local moments, if sufficiently strong, will allow magnetic ordering below a certain temperature, i.e. the Curie temperature in the present case.

## 4. ESR measurements of $\beta$-AgNpO$_2$(SeO$_3$)

Figure 6 shows the temperature dependent ESR spectra of $\beta$-AgNpO$_2$(SeO$_3$) when B//[001] (left) and B$\perp$[001] (right) for 44.89 GHz and 44.85 GHz, respectively. We observed a resonant absorption for both magnetic field orientations for sample 1, which was used for the ESR study. For B//[001], the absorption intensity increases at lower temperatures, which is typical for ESR. The ESR for B$\perp$[001] does not change significantly with temperature. Since the Np atoms (i.e J=4 for Np(V) and J= 9/2 for Np(IV)) have the largest magnetic moment in this material, it is clear that this ESR is due to the Np 5$f$ electrons. Although there are two Np sites per unit cell [6], only a single broad absorption line is observed. This suggests a non-negligible exchange interaction between the two Np sites [17], which is consistent with the magnetic ordering observed in the magnetization measurements.

From our analysis, the resonance field is proportional to the frequency and no gap is observed at zero magnetic field (not shown in the figures). This suggests that the ESR signal originates from a Kramers doublet, but this is not consistent with the magnetization studies which indicate the valence is Np(V) (i.e. a non-Kramers ion singlet ground state). The color of Np(IV) and Np(V) compounds are generally brown and green, respectively. Although the valence bond is 5+ from the chemical formula, the $\beta$-AgNpO$_2$(SeO$_3$) is brown in color which might suggests the existence of a finite concentration of Np(IV) (i.e. Kramers ion) [6]. We think that the minority concentration of Np(IV) is responsible for the ESR signal, since Np(V) is to a large extent ESR silent. The quantity of Np(IV) can be obtained from the comparison with a reference sample such as CuSO$_4$, or from Mossbauer, and this remains a subject for future investigation.

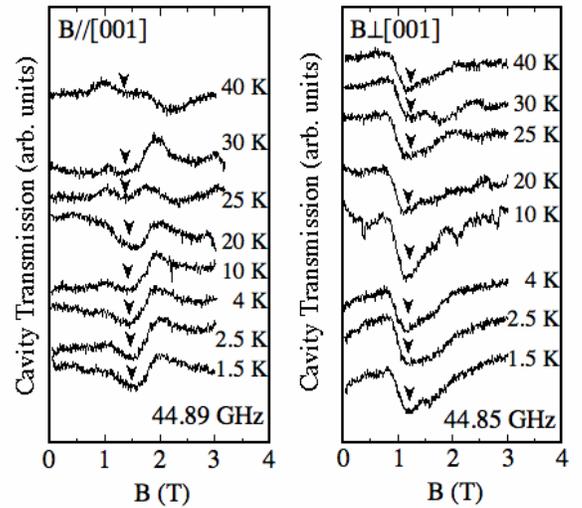

Figure 6. The temperature dependence of the ESR spectra of $\beta$-AgNpO$_2$(SeO$_3$) for magnetic field applied parallel (left) and perpendicular (right) to the [001] axis.

The *g*-value and the line width of the absorption line are shown in Fig. 7. The *g*-values do not show a significant change above 4 K (see Fig. 7(a)), and the values are 2.27±0.07 and 2.73±0.05 for $B//[001]$ and $B\perp[001]$, respectively. However, we observed a *g*-shift of 5% below 4 K for both axes, which is due to the evolution of the internal field created by the ferromagnetic ordering. The temperature, where the *g*-shift begins (i.e. 4 K), is lower than the $T_c$ of the other samples as obtained from the susceptibility measurements.

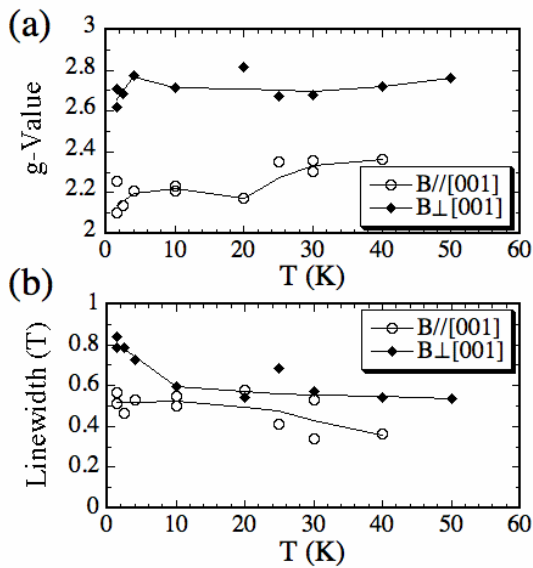

Figure 7. The temperature dependence of (a) the *g*-value, and (b) the line width for β-AgNpO$_2$(SeO$_3$). The circles and the diamonds show the results for B//[001] and B⊥[001], respectively.

Although the line width usually sharpens below the Curie temperature in a three-dimensional ferromagnetic system, the line width slightly broadens as the temperature decreases as shown in Fig. 7(b). This is possibly due to the two-dimensional sheets formed from cation-cation interactions in the structure of this system, or to an impurity effect since impurities with a rapid relaxation could cause an increase of the line width [18].

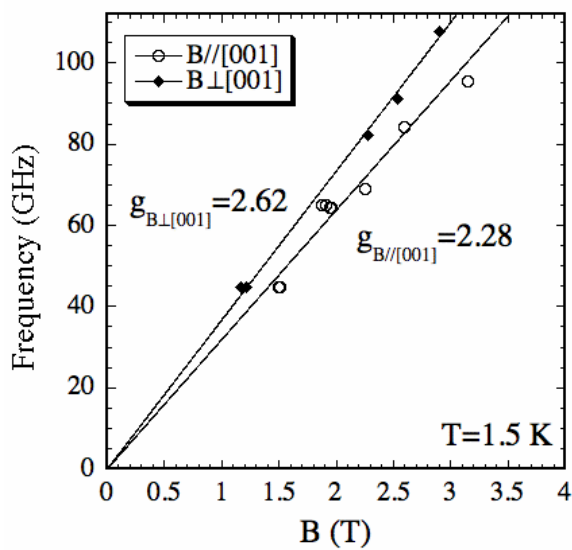

Figure 8. ESR frequency-resonance field diagram for β-AgNpO$_2$(SeO$_3$) at 1.5 K. The circles and the diamonds show the results for B//[001] and B⊥[001], respectively.

The ESR of a ferromagnet, i.e. ferromagnetic resonance (FMR), is well known and its resonance condition differs with sample shape [19]. In our case, the sample is small enough to treat as spherical. Therefore, there should be no contribution of the diamagnetic effect, and the resonance condition for FMR can be expressed as $h\nu=g\mu_B H$ (i.e. the same as for the EPR condition). Figure 8 shows the frequency-field diagram of the ESR line at 1.5 K (i.e. FMR). The observed FMR fits well with the resonance condition mentioned above, and its *g*-value is $g$=2.28 and $g$=2.62 for $B//[001]$ and $B\perp[001]$, respectively. The observation of FMR is consistent with the magnetization measurement.

## 5. Discussion and Conclusions

Our combined results from the magnetization, susceptibility and ESR measurements show that there is magnetic ordering in β-AgNpO$_2$(SeO$_3$) below a 8 K. Our analysis indicates the ordering must be ferromagnetic and that the valences of the neptunium in the compound are Np(IV) and Np(V). However, the relative proportion of Np(IV) is small since no hysteresis is observed in the magnetization cycles. In the ESR studies, the resonance arises from the Np(IV) sites which are influenced by magnetic order in both the Np(IV) and Np(V) populations. The resistance measurement shows that the electrons in this material are highly isolated, so that there is no contribution of conduction electrons to the magnetic moment. The overall value of the magnetic moment is smaller than the free ion value due to strong crystal field effects. The differences of the effective moments and $T_c$ values for the different samples are probably due to sample dependent effects, which may include the relative proportion of Np(IV) and Np(V).

**Acknowledgements**

This research was sponsored by the National Nuclear Security Administration under the Stewardship Science Academic Alliances program through DOE Research Grant #DE-FG03-03NA00066, and through NSF DMR 0203532. T.E.A.-S. acknowledges supported by the U.S. Department of Energy, Office of Basic Energy Sciences, Heavy Elements Program (Grant DE-FG02-01ER15187). We would like to thank P. Schlottmann for helpful discussions.